**Genome analyses and modelling the relationships between coding density, recombination rate and chromosome length**

Dorota Mackiewicz, Marta Zawierta, Wojciech Waga, Stanisław Cebrat

*Department of Genomics, Biotechnology Faculty, University of Wroclaw,*

*ul. Przybyszewskiego 63/77, 51-148 Wroclaw, Poland*

To whom correspondence should be addressed. E-mail: dorota@smorfland.uni.wroc.pl

**Abstract**

In the human genomes, recombination frequency between homologous chromosomes during meiosis is highly correlated with their physical length while it differs significantly when their coding density is considered. Furthermore, it has been observed that the recombination events are distributed unevenly along the chromosomes. We have found that many of such recombination properties can be predicted by computer simulations of population evolution based on the Monte Carlo methods. For example, these simulations have shown that the probability of acceptance of the recombination events by selection is higher at the ends of chromosomes and lower in their middle parts. The regions of high coding density are more prone to enter the strategy of haplotype complementation and to form clusters of genes which are "recombination deserts". The phenomenon of switching in-between the purifying selection and haplotype complementation has a phase transition character, and many relations between the effective population size, coding density, chromosome size and recombination frequency are those of the power law type.

**Key words**: crossover, complementation, gene linkage, Monte Carlo simulation.

**1. Introduction**

Mendel's law of independent assortment states that alleles placed at different loci assort independently of each other during the gamete formation. Actually, this is true only for genes that are not linked to each other which means that recombination frequency between them is very high. This law has been introduced into the neo-Darwinian theory by Fisher (1930) and Wright (1932) who made an explicit assumption that alleles are assorted independently into the gametes. Imagine a situation when several heterozygous loci located on one chromosome are considered. If one wants to follow the Fisher-Wright rule, the probability of separation of two alleles of neighbouring loci should be 0.5 and the total number of recombination events on the chromosome should be close to half of the number of the considered loci or higher. Notice that the recombination frequency depends on the number of considered heterozygous loci on chromosome. Fisher-Write model may render the proper results in description of some population evolution phenomena but only in those instances where the frequency of recombination does not play the critical role in the studies. It should not be used in cases where the recombination itself is the subject of studies like in the case of hybrydogenesis (Som and Reyer, 2006), or sexual reproduction of diploid organisms (Redfield, 1994). In the natural genomes, the series of genes are placed close to each other on the same chromosome and the probability of recombination between them is very limited (Daly et al., 2001; Greenwood et al., 2004; Hurst et al., 2004). Therefore the best representation of the Mendel law of independent assortment could be obtained for genes placed on different chromosomes. Nevertheless, it has been found that even in such configuration, clusters of human genes involved in the antigen recognition, which are located on different chromosomes are not inherited independently (Trowsdale, 2001; Hiby et al., 2004; Pharm, 2005; Gendzekhadze et al., 2006; Yawata et al., 2006) though, it is not established if it is a consequence of pre- or post-zygotic mechanisms. Analyses of recombination between homologous eukaryotic chromosomes revealed the uneven distribution of recombination events along the chromosomes. There are some regions where the probability of recombination is very low – recombination "deserts", and regions where the frequency of recombination is very high – recombination "hot spots" (Jeffreys et al., 2001; Petes, 2001; Yu et al., 2001; Mézard, 2006). Furthermore, there are some rules for distribution of recombination events along the chromosomes – the frequency of recombination is higher in the subtelomeric regions than in the central parts of chromosomes. The last statement is true for both, accepted recombination events in the real eukaryotic chromosomes (Kong et al., 2002; Nachman, 2002; Tenaillon et al 2002; Jensen-Seaman et al., 2004; Prachumwat et al. 2004; Barton et al., 2008) as well as for virtual chromosomes in the Monte Carlo computer simulations (Waga et al., 2007). In the computer models, the implemented recombination events were distributed evenly along the chromosomes but haploid gametes forming the surviving zygotes were produced by recombination at the ends of chromosomes rather than in the middle of chromosomes. Uneven distribution of recombination events along the chromosomes is related to the distribution of genes on it and can strongly influence the evolution of strings of the linked genes. Computer simulations have shown that depending on the internal and external genomic conditions the evolutionary costs could be lower for the strategy of complementing the functions of clustered (linked) genes or for the purifying selection, under high recombination rate. Switching between these two strategies of the haplotype evolution has a phase transition character (Zawierta et al., 2008) and it depends on the probability of recombination inside the cluster of genes (chromosome) and on the inbreeding coefficient (Waga et al., 2007). Inbreeding coefficient in natural populations depends on the effective size of populations ($N_e$) which usually are not very large and do not fulfil the rules governing the Mendelian population (ideal - infinite in size and panmictic). In models, inbreeding coefficient can be controlled by spatial distribution of individuals, restrictions in their relocations, mating restricted by distance between sexual partners and distances between parents and their offspring.

In this paper we compare the results of analyses of the virtual genetic structure of computer simulated populations with some structural properties of human chromosomes, especially the relations between the recombination rate, the coding density and a strategy of their evolution.

**2. Materials and methods**

*2.1. Statistics of human chromosomes and recombinations*

The size of human chromosomes, numbers of genes and their distribution were downloaded from the Ensembl Web site (http://www.ensembl.org/Homo_sapiens) (Release 56). Genetic maps of human chromosomes and localizations of genetic markers were downloaded from HapMap Web site (The International HapMap Consortium, 2005). To compare the results of computer simulations with the human genome data we have used the physical size of human chromosomes being the distance between the first and the last mapped markers on the genetic map of chromosome, measured in nucleotides.

*2.2. The model*

In our simulations, each Monte Carlo step (MCs) starts with a fixed-sized population of N individuals. Each individual is represented by its diploid genome composed of two homologous chromosomes (bitstrings L bits long). Bits correspond to genes, so if a bit is set to 0 – it corresponds to the functional (wild) allele. However, if a bit is set to 1 – it corresponds to the defective version of an allele. All defective alleles are recessive which means that both alleles at the corresponding positions in the two bitstrings (loci) have to be defective to determine the deleterious phenotype. The genomes of all newborns are checked and those with at least one defective phenotype are killed. Additionally, during each MCs 2% of populations is randomly killed. To fill up the gap, the genomes of newborns are constructed in the reproduction process where two individuals are randomly drawn from the survivors' pool as the mating partners. Genome of each partner is replicated and one mutation is introduced into each copy of the bitstring in the random position. If the value of a bit at position chosen for mutation is 0 – it is changed to 1 if it is 1 – it stays 1 (there are no reversions). The two copies of bitstrings recombine with probability C in the process imitating crossover at the same, randomly chosen positions. One copy of bitstring or one product of recombination, if crossover has occurred, is considered to be a gamete. The haploid gametes of two partners fuse and form the diploid zygote.

New individuals are being produced until the population size reaches N. The varying parameters of the model are:

N – population size,

L – number of bits in the bitstring,

C – recombination rate – the probability of recombination between bitstrings.

To show the differences between the two haplotypes of individuals we have analyzed the Hamming distance between them which is a sum of all corresponding loci which have different values of bits (0,1 or 1,0). In this case the Hamming distance represents the number of heterozygous loci.

## 3. Results and discussion

*3.1. Complementary haplotype strategy versus purifying Darwinian selection*

In simulations performed according to the described model, it has been found that populations evolving under the constant parameters of N and L may choose between two strategies of their genome evolution depending on the recombination rate C, namely – purifying Darwinian selection with relatively low fraction of defective alleles or complementation of large fragments of bitstrings (Waga et al., 2007; Zawierta et al., 2007). If recombination is switched off (C=0), after simulations being long enough, the populations reach a final stage with perfectly matched bitstrings – all loci are heterozygous. Switching between these two strategies has a very sharp transition character from purifying to the complementation strategy at critical recombination frequency. In panmictic populations the critical value of recombination depends on N. In fact, it depends on effective population size or inbreeding coefficient. When simulations are performed with large populations but with mating spatially restricted by distance between partners – the critical value of recombination does not depend on the value of N. Though, the phenomenon has properties of phase transition (Zawierta et al., 2008).

The results of series of simulations for two different N and constant L are shown in Fig. 1. Under low recombination rate the complementing strategy is winning which results in high Hamming distance between bitstrings, corresponding to the high number of heterozygous loci. Under higher recombination rate the purifying selection strategy, with lower fraction of heterozygous loci is winning. In larger populations, the transition from complementation to purifying selection happens at a lower recombination rate i.e. for N = 1000 the transition occurs between C = 0.18 and 0.19, for smaller populations, the transition is observed at a higher recombination rate (for N = 100 the transitions occurs between C = 0.7 and 0.71). The recombination rate at the transition will be called critical value and denominated by $C^*$. What is also interesting is the fact that close to the transition point the ends of chromosomes are under purifying selection while the central parts of chromosome are rather complement. This observation suggests that uneven distribution of recombination events along the chromosomes should be observed – the higher recombination rates in the subtelomeric regions and the lower recombination rates in the central parts of chromosomes which agrees with the data obtained from the experimental studies of recombination in different organisms (Kong et al., 2002; Nachman, 2002; Tenaillon et al 2002; Jensen-Seaman et al., 2004; Prachumwat et al. 2004; Barton et al., 2008).

*3.2. Chromosome length and coding density*

In models, the chromosome size is described by L – the number of bits in the bitstring, corresponding to the number of genes. The length of natural chromosomes can be measured in three different ways (units): number of genes, physical units corresponding directly to the number of nucleotide pairs, and genetic units – centiMorgans corresponding to the frequency of recombinations (one centiMorgan is a fragment of chromosome where the probability of crossover is 1 percent). The physical size of human chromosomes ranges from 35.1 Mb (Mb – million base pairs) to the 247.1 Mb (The International HapMap Consortium, 2005) and strongly correlates with genetic size measured in centiMorgans (Fig. 2 A). The number of genes contained in the human autosomes ranges from a few hundreds to more than two thousands. What is more important, some chromosomes of approximately the same size measured in base pairs or in centiMorgans contain significantly different number of genes (i.e. chromosomes 17, 18, 19 – see Table 1). As a result, the correlation between the size of chromosomes (both, genetic and physical) and their coding capacity measured in number of genes is not high (Fig. 2 B and C). Moreover, human chromosomes differ substantially in coding density measured in number of genes per recombination unit and coding density is not correlated with the chromosome size (Fig. 2 D). The highest coding density in the human genome is observed for the chromosome 19 and it is more than 6 times higher than for the smallest coding density observed for the chromosome 18. It doesn't matter if we consider the size of chromosomes measured in base pairs or recombination units (Venter et al., 2001; Grimwood et al., 2004; Nusbaum et al., 2005). The length of the chromosome 19 is almost four times smaller than that of the chromosome 2 which has a similar number of genes. What is interesting, almost thirty percent of genes located on the chromosome 19 are members of large, clustered gene families. In addition, majority of these familial clusters are represented by syntenically homologous clusters in mouse (Dehal et al., 2001). Therefore, to find out how the strategy of chromosome evolution depends on the coding density seems to be very important.

*3.3. Chromosome evolution*

In the computer models, the number of bits in the bitstrings (L) corresponding to the number of genes in chromosomes is usually used as synonym of their size. Recombination frequency (C) is not considered as a phenomenon concerning the chromosomal size. In fact, it determines the size of chromosomes in the genetic units measured in centiMorgans. If only one of these two parameters is changed – L or C, the coding density – measured in number of genes per centiMorgan is also changed. To show the effect of coding density on the evolution of genomes we have simulated populations with different size of bitstrings (L) and different recombination rate (C) and, looked for the C* value for different population size. Results are shown in Figures 3 and 4. Notice that both figures are in log/log scale.

The relation between critical recombination rate per bitstring and the number of bits in the bitstring for different population size is presented in Fig. 3. Bitstrings containing more bits (genes) enter the complementing strategy under higher recombination rate than those containing fewer bits, i.e. bitstrings 2048 bits long enter the complementing strategy at 2.5 crossovers per gamete production on average in the efficient population size of the order of 100 individuals. Those data correspond to the human chromosome 1, which contains above 2100 genes and is about 278 centiMorgans long.

In order to show more clearly how the critical recombination rate depends on the population size, Fig. 3 has been transposed into Fig. 4, where the relation between critical recombination rate and the effective population size is shown for bitstrings containing different number of bits. Since both plots are linear in the log/log scale it means that the relations are of the power law type. For longer bitstrings the approximately parallel lines are shifted up. The results suggest that the chromosomes containing more genes enter the complementing strategy more easily. It agrees with the relation between the size of chromosomes and the values of the linkage disequilibrium (LD) found in natural genomes. It has been found that large chromosomes have generally stronger linkage disequilibrium in comparison to small chromosomes (Smith et al., 2005). One should remember that the human chromosomes differ significantly in their coding density measured as number of genes by centiMorgan and it is not correlated with the size of chromosomes.

The relation between the numbers of genes per centiMorgan at critical recombination rate C* and the numbers of genes in the bitstrings is presented in Fig. 5. Density of genes per centiMorgan at C* is growing non-linearly with the number of genes in the bitstrings if the efficient population size is constant. It can be expected that if the crossover frequency is proportional to the number of genes, the longer chromosomes should enter the complementing strategy easier. In addition, we have counted the number of genes per centiMorgan in human chromosomes and placed the corresponding values in the space of plots shown in Fig. 5. The majority of these values are located in the area of critical recombination rates characteristic for effective population size between 100 – 200 individuals, with the exception of chromosome 19, which is located between the line for population size of 300 and 400. Furthermore, it is worth noticing that the effective size of our virtual populations corresponds to the much higher real effective size of populations because in our model all individuals are in the reproduction age while in the human (age structured) populations approximately only one third of individuals is in the reproduction age (Cavalli-Sforza et al., 1994). Therefore, the effective size of virtual populations should be multiplied by three, to compare with natural populations. The effective size of populations ($N_e$) for critical value of recombination which was suggested by computer modelling seems to be lower than the usually quoted value of 10 000 (Takahata, 1993). However, Liu et al. (2006) after incorporating the geographic distances between populations, estimated an $N_e$ value for about 1000 individuals for the founding population. Thus, it seems reasonable to assume that humans evolved in the populations small enough to put the chromosomes close to the critical values of recombination. If the last suggestions are correct, the recombination events along the human chromosomes should be distributed unevenly, showing higher recombination rate in the subtelomeric regions than in the chromosome centres, like it has been shown previously in the computer modelling (Waga et al, 2007). In fact, such an uneven distribution of recombination events has been observed for human chromosomes (Payseur and Nachman, 2000; Kong et al., 2002; Nachman, 2002; Jensen-Seaman et al., 2004; Cheung et al., 2007). As a result, on average the LD is lower near telomeres than near centromers (The International HapMap Consortium, 2005). It is agreed with data obtained from the detailed studies of linkage disequilibrium chromosome 19, where it was found several regions of excessively high LD near the centromere (Phillips et al., 2003).

In Fig. 6 we have compared the real data obtained for human chromosomes with our virtual chromosomes. We have used the detrended cumulative plots to present the distribution of accepted recombination events in two human chromosomes and in one virtual chromosome after simulations. X-axis represents position on chromosome normalized to 1, while y-axis represents the normalized, cumulated differences between the found recombination frequencies and the recombination expected to occur under assumption that it is distributed evenly. Thus, the increasing parts of the plots correspond to the relatively high recombination rate while decreasing parts correspond to the regions of chromosomes where recombination rate is lower than average for this chromosome. The same characteristic distributions of recombination events can be observed for all human autosomes. The similarity between results of simulations and experimental data from human chromosomes suggests that human chromosomes have evolved close to the critical values of recombination. Since the critical values of recombination are closely related to the effective population sizes, it could be concluded that humans have evolved in the relatively small populations. Recent results of simulations, performed with a model with many possible morphs of each gene confirm the results shown in this paper for the model assuming only two alleles for each gene (Cebrat et al, 2009).

## 4. Conclusions

Computer simulations of population evolution predict the uneven distribution of accepted recombination events inside the chromosomes. The higher recombination rates in the subtelomeric regions and the lower recombination rates in their central parts mean that these chromosomes recombine with frequency close to its critical value, where the switching between the two strategies of population evolution – the complementing haplotypes and purifying selection is possible. The switching between those strategies depends on the chromosome length, their coding density, recombination rate and inbreeding. Regions with high coding density are more prone to enter the gene clustering strategy. Comparisons of the properties of the human genomes with the virtual genomes suggests that humans evolved in relatively small effective populations.

## 5. Acknowledgements

The work has been done in the frame of European programs: COST Action MP0801, FP6 NEST - GIACS and UNESCO Chair of Interdisciplinary Studies, University of Wroclaw. Calculations have been carried out in Wroclaw Centre for Networking and Supercomputing (http://www.wcss.wroc.pl), grant #102.

**6. References**

Barton, A. B., Pekosz, M. R., Kurvathi, R. S., Kaback, D. B., 2008. Meiotic recombination at the ends of chromosomes in *Saccharomyces cerevisiae*. Genetics 179, 1221-1235.

Cavalli-Sforza, L. L, Menozzi, P., Piazza, A., 1994. The history and geography of human genes. Princeton University Press, Princeton, NJ.

Cebrat, S., Stauffer, D., Sa Martins, J. S., Moss de Oliveira, S., de Oliveira, P. M. C., 2009. Modelling survival and allele complementation in the evolution of genomes with polymorphic loci. arXiv:0911.0589.

Cheung, V. G., Burdick, J. T., Hirschmann, D., Morley, M., 2007. Polymorphic Variation in Human Meiotic Recombination. Am. J. Hum. Genet. 80, 526-30.

Daly, M. J., Rioux, J. D., Schaffner, S. F., Hudson, T. J., Lander, E. S., 2001. High-resolution haplotype structure in the human genome. Nat. Genet. 29, 229-232.

Dehal P, Predki P, Olsen AS, Kobayashi A, Folta P, Lucas S, Land M, Terry A, Ecale Zhou CL, Rash S, et al., 2001. Human chromosome 19 and related regions in mouse: conservative and lineage-specific evolution. Science. 293, 104-111.

Fisher, R. A., 1930. *The genetical theory of natural selection*. Clarendon Press, Oxford.

Gendzekhadze, K., Norman, P. J., Abi-Rached, L., Layrisse, Z., Parham, P., 2006. High KIR diversity in Amerindians is maintained using few gene-content haplotypes. Immunogenet. 58, 474-480.

Greenwood, T. A., Rana, B. K., Schork N. J., 2004. Human haplotype block sizes are negatively correlated with recombination rates. Genome Res. 14, 1358-1361.

Grimwood, J., Gordon, L. A., Olsen, A., Terry, A., Schmutz, J., Lamerdin, J., Hellsten, U., Goodstein, D., Couronne, O., Tran-Gyamfi, M., et al., 2004. The DNA sequence and biology of human chromosome 19. Nature 428, 529-35.

Hiby, S. E., Walker, J.J., O'shaughnessy, K. M., Redman, C.W., Carrington, M., Trowsdale, J., Moffett, A. J., 2004. Combinations of maternal KIR and fetal HLA-C genes influence the risk of preeclampsia and reproductive success. Exp. Med. 200, 957-965.

Hurst, L. D., Pal, C., Lercher, M. J., 2004. The evolutionary dynamics of eukaryotic gene order. Nature Rev. Genet. 5, 299-310.

Jeffreys, A. J., Kauppi, L., Neumann, R., 2001. Intensely punctate meiotic recombination in the class II region of the major histocompatibility complex. Nat. Genet. 29, 217-222.

Jensen-Seaman, M. I., Furey, T. S., Payseur, B. A., Lu, Y., Roskin, K. M., Chen, C.F., Thomas, M.A., Haussler, D., Jacob, H.J., 2004. Comparative recombination rates in the rat, mouse, and human genomes. Genome Res. 14, 528-538.

Kong, A., Gudbjartsson, D. F., Sainz, J., Jonsdottir, G. M., Gudjonsson, S. A., Richardsson, B., Sigurdardottir, S., Barnard, J., Hallbeck, B., Masson, G., Shlien, A., Palsson, S. T., Frigge, M. L., Thorgeirsson, T. E., Gulcher, J. R., Stefansson, K., 2002. A high-resolution recombination map of the human genome. Nat. Genet. 31, 241-247.

Liu, H., Prugnolle, F., Manica, A., and Balloux, F. 2006. A geographically explicit genetic model of worldwide human-settlement history. Am. J. Hum. Genet. 79: 230-237.

Mézard, C., 2006. Meiotic recombination hotspots in plants. Biochem Soc. Trans. 34, 531-534.

Nachman, M.W., 2002. Variation in recombination rate across the genome: evidence and implications. Curr. Op. Genet. Devel. 12, 657-663.

Nusbaum, C., Zody, M. C., Borowsky, M. L., Kamal, M., Kodira, C. D., Taylor, T. D., Whittaker, C. A., Chang, J. L., Cuomo, C. A., Dewar, K., et al., 2005. DNA sequence and analysis of human chromosome 18. Nature 437, 551-555.

Parham, P., 2005. MHC class I molecules and KIRs in human history, health and survival. Nat. Rev. Immunol. 5, 201-214.

Payseur, B. A., Nachman, M. W. 2000. Microsatellite variation and recombination rate in the human genome. Genetics 156, 1285-1298.

Petes, T. D., 2001. Meiotic recombination hot spots and cold spots. Nat. Rev. Genet. 2, 360-369.

Phillips, M. S., Lawrence, R., Sachidanandam, R., Morris, A. P., Balding, D. J., Donaldson, M. A., Studebaker, J. F., Ankener, W. M., Alfisi, S. V., Kuo, F. S., et al. 2003. Chromosome-wide distribution of haplotype blocks and the role of recombination hot spots. Nat. Genet. 33, 382-387.

Prachumwat, A., DeVincentis, L., Palopoli, M. F., 2004. Intron size correlates positively with recombination rate in *Caenorhabditis elegans*. Genetics 166, 1585-1590.

Redfield, R. J., 1994. Male mutation rates and the cost of sex for females. Nature 369, 145–147.

Smith, A. V., Thomas, D. J., Munro, H. M., Abecasis, G. R., 2005. Sequence features in regions of weak and strong linkage disequilibrium. Genome Res.15, 1519-34.

Som, C., Reyer, H-U. 2006. Variation in sex ratio and evolutionary rate of hemiclonal *Rana esculenta* populations. Evol. Ecol. 20, 159-172.

Takahata, N., 1993. Allelic genealogy and human evolution. Mol. Biol. Evol. 10, 2-22.

Tenaillon, M. I., Sawkins, M. C., Anderson, L. K., Stack, S. M., Doebley, J., Gaut, B. S., 2002. Patterns of diversity and recombination along chromosome 1 of maize (Zea mays spp. mays L.). Genetics 162, 1401–1413.

The International HapMap Consortium, 2005. A haplotype map of the human genome. Nature 437, 1299–1320.

Trowsdale, J., 2001. Genetic and functional relationships between MHC and NK receptor genes. Immunity 15, 363-374.

Venter, J. C., Adams, M. D., Myers, E. W., Li, P. W., Mural, R. J., Sutton, G. G., Smith, H. O., Yandell, M., Evans, C. A., Holt, R. A., et al., 2001. The sequence of the human genome. Science 291, 1304-1351.

Waga, W., Mackiewicz, D., Zawierta, M., Cebrat, S., 2007. Sympatric speciation as intrinsic property of expanding populations. Theory Biosci. 126, 53 - 59.

Wright, S., 1932. The roles of mutation, inbreeding, crossbreeding and selection in evolution. Proceedings of the 6th International Congress of Genetics. 1, 356-366.

Wright, S., 1931. Evolution in Mendelian populations. Genetics 16, 97-159.

Yawata, M., Yawata, N., Draghi, M., Little, A.M., Partheniou, F., Parham, P., 2006. Roles for HLA and KIR polymorphisms in natural killer cell repertoire selection and modulation of effector function. J. Exp. Med. 203, 633-645.

Yu, A., Zhao, C., Fan, Y., Jang, W., Mungall, A. J., Deloukas, P., Olsen, A., Doggett, N. A., Ghebranious, N., Broman, K. W., Weber, J. L., 2001. Comparison of human genetic and sequence-based physical maps. Nature 409, 951-953.

Zawierta, M., Biecek, P., Waga, W., Cebrat, S. 2007. The role of intragenomic recombination rate in the evolution of population's genetic pool. Theory Biosci. 125, 123-132.,

Zawierta, M., Waga, W., Mackiewicz, D., Biecek, P., Cebrat, S. 2008. Phase Transition in Sexual Reproduction and Biological Evolution. Int. J. Mod. Phys. C19, 917-926.

**Table 1**

Number of protein coding genes on each human autosome.

| | Chromosome | | | Number of genes[a] |
|---|---|---|---|---|
| Nr | Assembly size (Mb)[a] | Physical length (Mb)[b] | Genetic length (cM)[b] | |
| 1 | 249.3 | 247.1 | 278.1 | 2163 |
| 2 | 243.2 | 242.7 | 263.4 | 1390 |
| 3 | 198.0 | 199.3 | 224.6 | 1168 |
| 4 | 191.2 | 191.3 | 213.2 | 911 |
| 5 | 180.9 | 180.6 | 204.0 | 993 |
| 6 | 171.1 | 170.8 | 193.0 | 1110 |
| 7 | 159.1 | 158.7 | 187.0 | 1115 |
| 8 | 146.4 | 146.1 | 170.2 | 839 |
| 9 | 141.2 | 140.2 | 168.3 | 889 |
| 10 | 135.5 | 135.3 | 179.4 | 876 |
| 11 | 135.0 | 134.3 | 159.5 | 1445 |
| 12 | 133.9 | 132.3 | 173.0 | 1148 |
| 13 | 115.2 | 96.2 | 127.2 | 369 |
| 14 | 107.3 | 87.7 | 117.1 | 707 |
| 15 | 102.5 | 82.0 | 131.4 | 753 |
| 16 | 90.4 | 88.7 | 135.0 | 1042 |
| 17 | 81.2 | 78.6 | 129.5 | 1308 |
| 18 | 78.1 | 76.1 | 119.6 | 317 |
| 19 | 59.1 | 63.6 | 107.9 | 1544 |
| 20 | 63.0 | 62.4 | 108.1 | 608 |
| 21 | 48.1 | 37.0 | 62.3 | 280 |
| 22 | 51.3 | 35.1 | 73.6 | 529 |

[a]Assembly size human chromosomes means the size of the chromosome based on the complete genome assembly.

[b]The physical size is the distance in Mb between the first and last mapped markers and genetic length is the distance in cM between the first and last mapped markers.

**Figures Legends:**

Fig. 1. Fraction of heterozygous loci along the virtual chromosomes in populations of different sizes [N] and recombination rates [C]. (A) N = 100 individuals, (B) N = 1000 individuals. Notice, for larger populations complementing regions appear under lower recombination rate.

Fig. 2. (A) The relation between the physical length of human chromosomes in millions base pairs and their length in genetic units – centiMorgans. (B) The relation between the genetic length of human chromosomes in centiMorgans and coding capacity measured in the number of genes per chromosome. (C) The relation between the physical length of human chromosomes in millions base pairs and their coding capacity measured in the number of genes per chromosome. (D) The relation between the coding density and the chromosome size.

Fig. 3. Critical recombination rate [C*] as a function of length of virtual chromosomes measured in their coding capacity (number of genes) for different efficient population sizes [N]. See text for details.

Fig. 4. Critical recombination rate for different length of chromosomes [L] in the relation with different population sizes. Notice the power law relation between the recombination rate and efficient population size – both axes in logarithmic scale.

Fig. 5. Density of genes per centiMorgan at critical recombination rate for chromosomes containing different numbers of genes and for different population sizes

[N]. Under conditions above the lines chromosomes choose the complementing strategy for a given population size while below the lines – the purifying strategy. Notice that short chromosomes can choose the complementing strategy at lower coding density per recombination unit. Solid circles mark the positions of human chromosomes in that virtual space (number of genes per centiMorgan versus total number of genes in chromosome). Under those conditions, some human chromosomes could choose the complementing strategy in the efficient population size of the order of 300 individuals at the reproduction age.

Fig. 6. The detrended cumulative plot of the distribution of accepted recombinations along the chromosomes. Dense points (look like a bold line) – human chromosomes 13 and 20, x – virtual chromosome of individuals evolving close to critical value of recombination. Increasing parts of plots represent the higher recombination rate than the average, decreasing parts show the regions where the recombination rate is lower than the average.

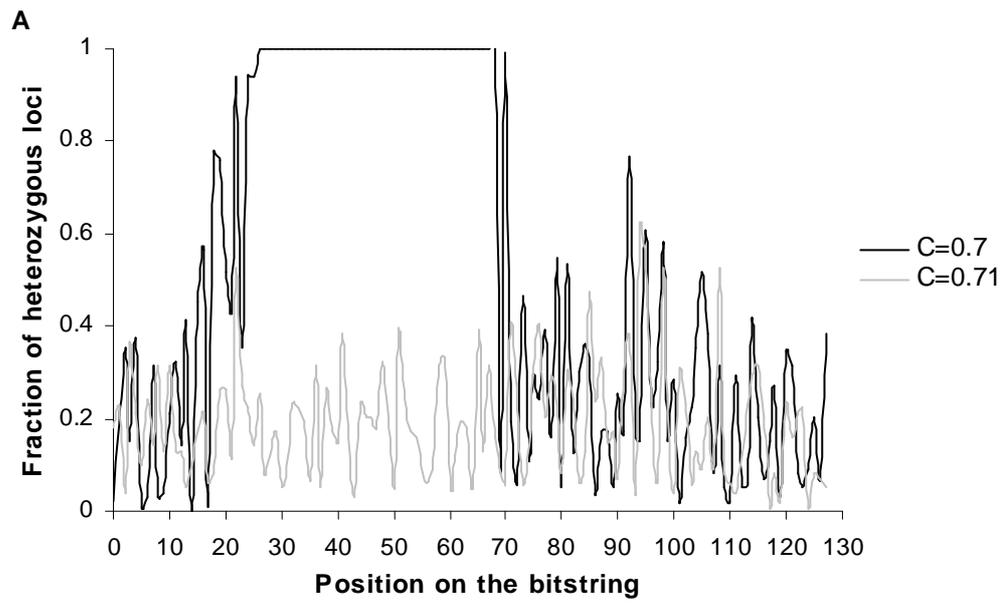

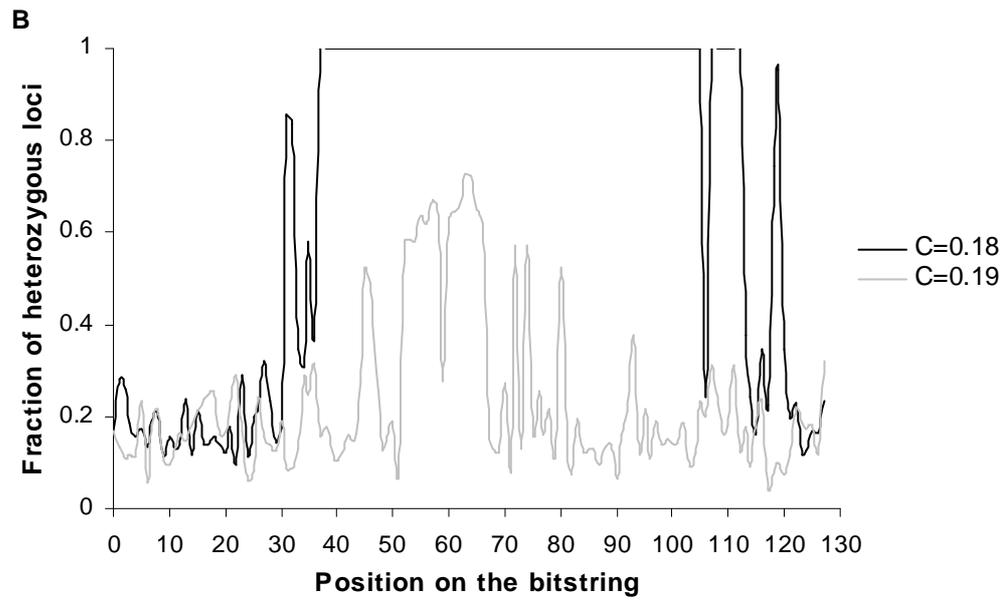

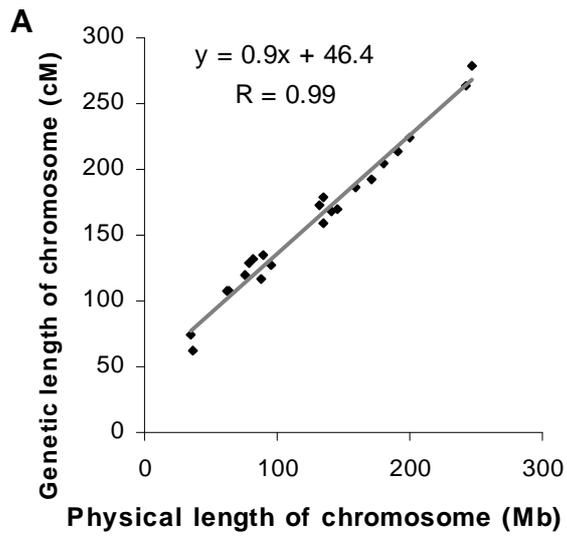
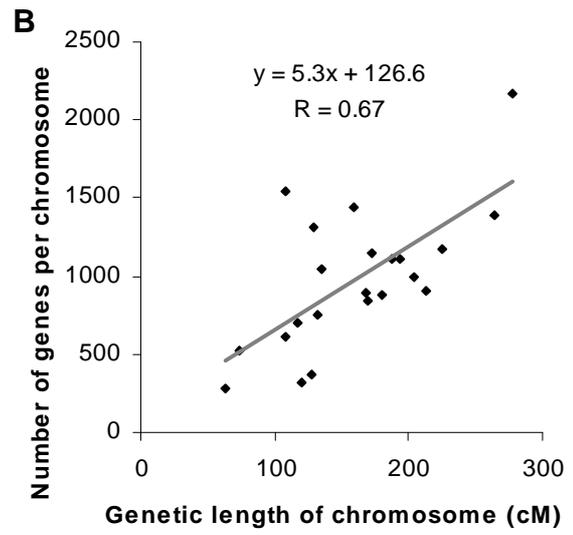
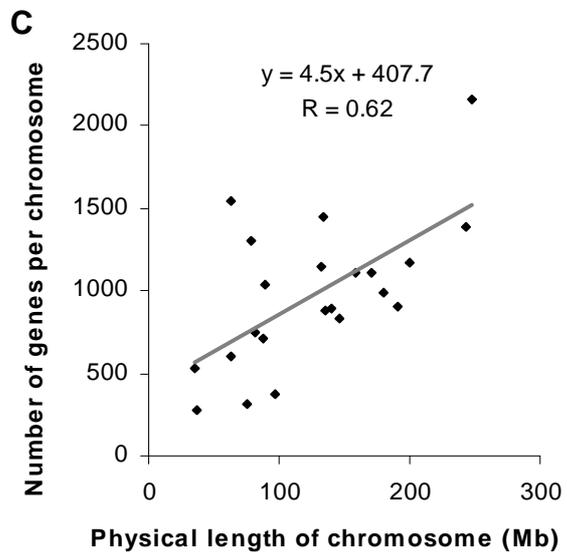
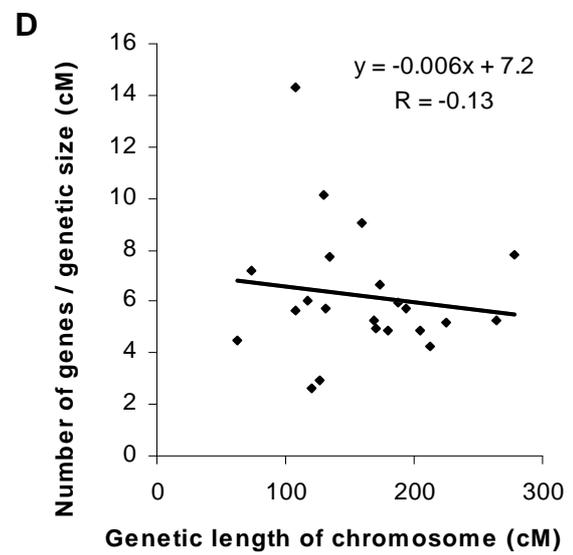

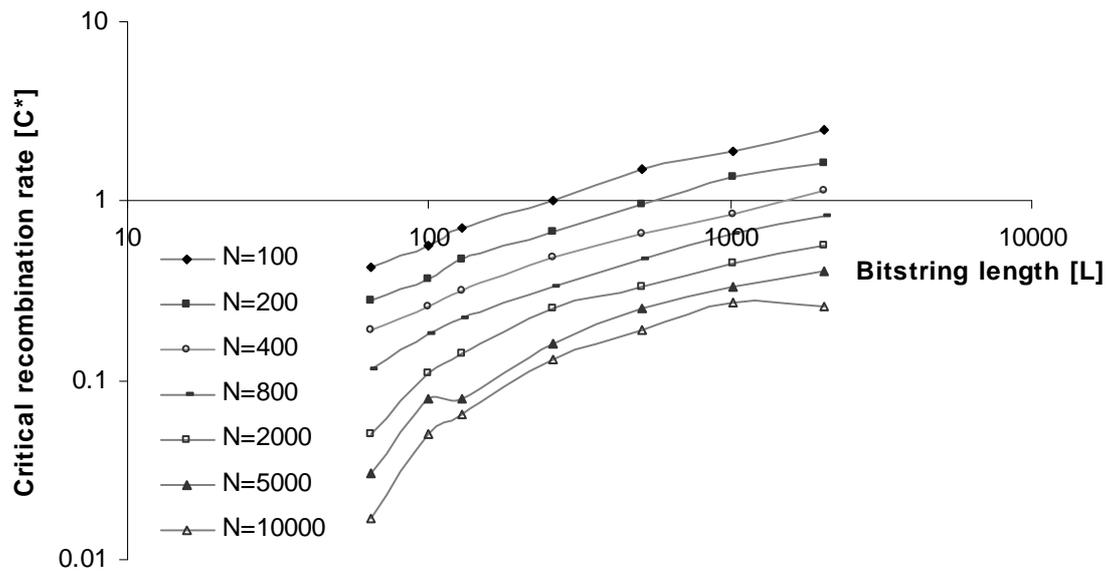

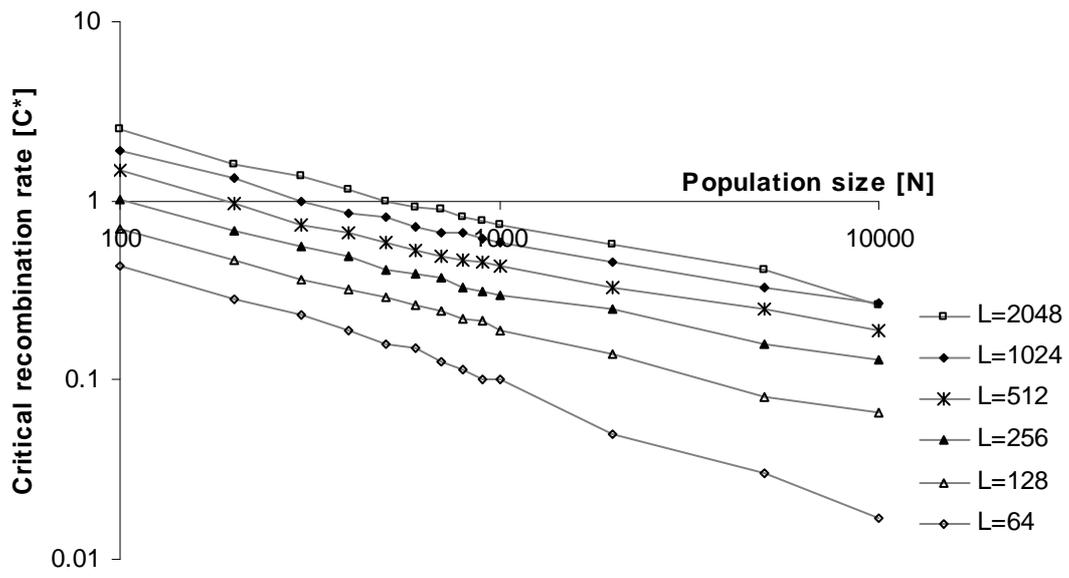

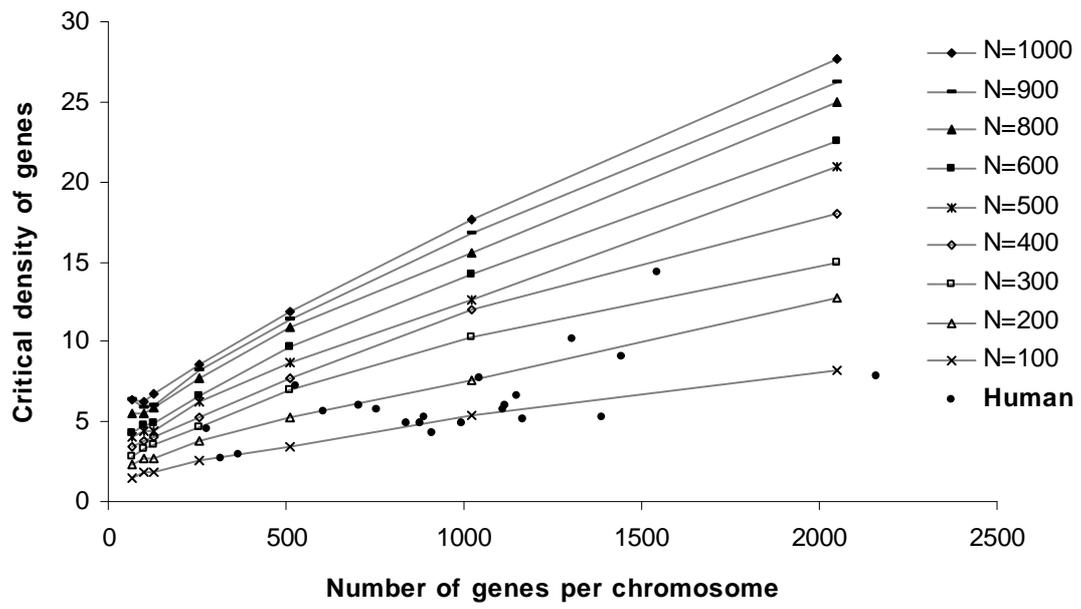

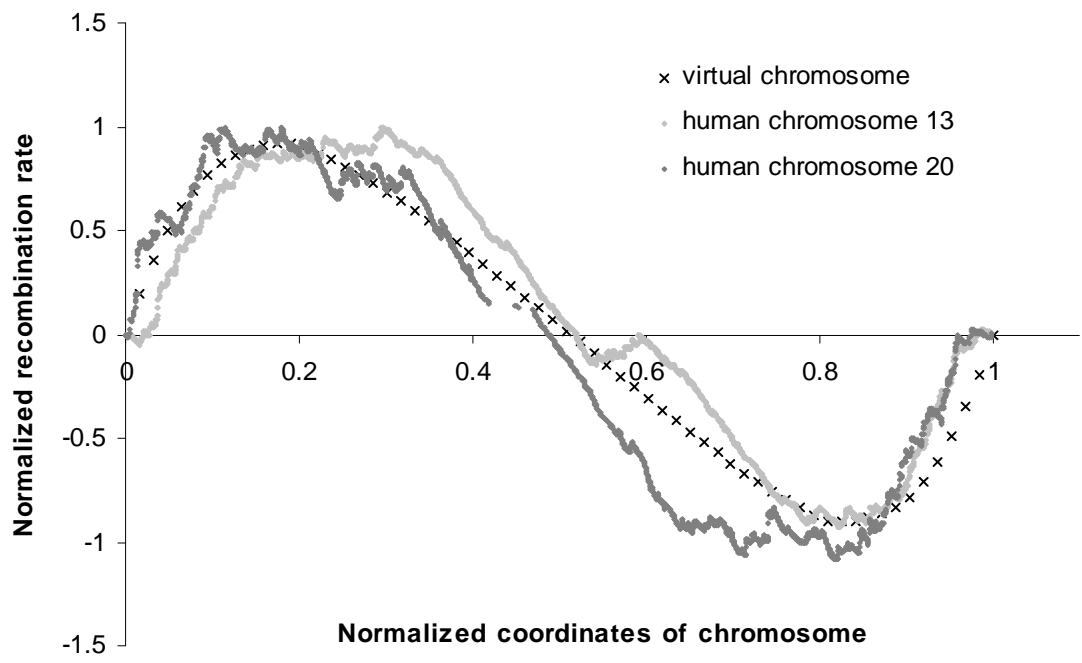